\documentclass[a4paper,11pt]{article}

\usepackage{jheppub} 
\usepackage[T1]{fontenc} 
\usepackage{hyperref}
\usepackage{bbold}
\usepackage{xcolor}
\usepackage{tikz}
\usepackage{amsfonts}
\def\beq{\begin{equation}}
\def\eeq{\end{equation}}
\def\beqn{\begin{eqnarray}}
\def\eeqn{\end{eqnarray}}
\def\bpm{\begin{pmatrix}}
\def\epm{\end{pmatrix}}

\def\half{\tfrac{1}{2}}
\def\nat{\natural}
\def\sA{{\scriptscriptstyle A}}
\def\sB{{\scriptscriptstyle B}}
\def\sC{{\scriptscriptstyle C}}
\def\sD{{\scriptscriptstyle D}}

\title{\boldmath  Some Hidden Structure in BKL.}

\author[a,b,c]{Malcolm J. Perry }
\affiliation[a]{School of Physics and Astronomy, Queen Mary University of London, Mile End Road, London E1 4NS, UK.}
\affiliation[b]{DAMTP, Centre for Mathematical Sciences, Wilberforce Road, Cambridge, CB3 0WA, UK.}
\affiliation[c]{Trinity College, Cambridge, CB2 1TQ, UK.}
\emailAdd{malcolm@maths.cam.ac.uk}

\abstract{The way spacetime behaves as one approaches a spacelike singularity is re-investigated. 
We find a simple twistorial presentation that includes and simplifies the classic work of
Belinskii, Khalatnikov and Lifshitz as well as the more recent results of Damour, Henneaux and Nicolai. 
We speculate on the application of our technique to the $E_{10}$
programme of M-theory.  
}

\begin{document}

\maketitle
\flushbottom

\section{Introduction}
\label{sec:intro}
Classical general relativity leads to  the formation of spacetime singularities under various circumstances.
The singularity theorems of Penrose \cite{Penrose:1964wq} and Penrose and 
Hawking \cite{Hawking:1970zqf} show that timelike or null geodesics terminate
after a finite distance. In numerous examples, singularities are found to be regions of infinite 
curvature {\footnote{For example in black holes or the big bang and big crunch in cosmology.}}.  
The fact that singularities are inevitable and lead to a complete lack of predictability 
led  John Wheeler \cite{Wheeler:Woolf} to write that\, \lq\lq Einstein's general 
relativity gives not the slightest evidence whatsoever for a \lq before\rq\ before the 
big bang or an \lq after\rq\ after collapse.\rq\rq\ Singularities are best interpreted 
as being the boundaries of spacetime 
\cite{Schmidt:1971uf}.

Investigations into the geometry of spacetime close to spacelike singularities was inspired by Landau and 
eventually carried out to completion 
by Belinskii, Khalatnikov and E.~M.~Lifshitz (BKL) \cite{Belinsky:1970ew}.
{\footnote{See also the much earlier and incomplete work of Khalatnikov and 
E.~M.~Lifshitz. \cite{Lifshitz:1963ps}}
This work was subsequently revisited by Damour, Henneaux, Julia and Nicolai 
\cite{Damour:2001sa}, Damour, Henneaux and Nicolai (DHN) \cite{Damour:2002et} 
and it is this latter treatment that 
we follow in this paper. Useful reviews are provided by the paper of 
Henneaux, Persson and Spindel \cite{Henneaux:2007ej} and the
book of Belinskii
and Henneaux \cite{Belinski:2017fas}. 
The essential result of BKL is that as one approaches a spacelike singularity, points in space
become causally separated and therefore behave independently of each other.
In this paper we are mainly concerned with four spacetime dimensions. 
Both BKL and DHN show that as one approaches a spacelike singularity, 
the time evolution of each point in space is that of a Bianchi I universe but
interrupted by rapid changes in both the rate of expansion or contraction of the principal directions and
by rotation of the axes of the principal directions. DHN
showed that the minisuperspace of a Bianchi I universes is  three-dimensional flat Minkowski space. 
Evolution is then described by a null geodesic in this minisuperspace with the
\lq\lq interruptions\rq\rq\ of BKL corresponding to the null geodesic being specularly reflected 
off a collection of walls in the minisuperspace. 

DHN \cite{Damour:2002cu} and also Damour and Nicolai \cite{Damour:2007bd} then conjectured 
that the structure revealed by BKL 
sheds light on the fundamental degrees of freedom of the gravitational field.
Their observation was specific to the case of M-theory where they speculated
that the Kac-Moody algebra $E_{10}$ controlled the complete theory. BKL showed that
points in space became independent of each other as one approached a spacelike singularity. 
DHN found some evidence leading to the opposite; 
by considering an $E_{10}$ coset model, they found indications that space emerged 
from a more fundamental structure. 
Finding the fundamental degrees of freedom of any theory is historically not so straightforward. For the 
case of weak interactions, it was not until it was realised that the W and Z bosons would allow
for a renormalizable theory that one could claim an understanding. For the strong interactions, it
was not until the discovery of quarks, color and the renormalizability of QCD that one had a 
coherent picture. For the case of gravity, it seems that one is still in the dark since although
string theory provides a finite theory that includes perturbative gravitation, and M-theory goes
somewhat beyond, neither is a quantum theory of spacetime. DHN
found the $E_{10}$ approach was extremely promising but their picture is currently incomplete.
One therefore wonders
if there is a simple modification that would lead to further progress with the stimulating ideas
of DHN and their collaborators.  

In section two, we briefly outline the picture provided by BKL in dimension four. The BKL 
result is best understood by seeing that the bouncing Bianchi I spacetimes
are piecewise null geodesics in a minisuperspace. The minisuperspace turns out to be 
just three-dimensional Minkowski spacetime together with the walls that produce the bounces. 
In section three,
we provide a spinorial description of the null geodesics. 
The principal result here is that we find a 
description of the phase space of the BKL picture rather than just the configuration space. 
The approach is based on  the special properties of $SL(2,\mathbb{R})$ and so is restricted 
to examining the case of four spacetime dimensions. In section four, we show that it seems more natural
to formulate our description in terms of twistors and the conformal group of the minisuperspace, $SO(3,2)$
or its double cover
$Sp(4,\mathbb{R})$. \footnote{For clarity, we note here our use of various indices. $i,j,k \ldots$
run over $1,2,3$ and are spatial indices in spacetime. $a,b,c \ldots$ are minisuperspace indices and run 
over $0,1,2$.
$A,B,C \ldots$ are spinor indices of $SO(2,1)$ and can be $0$ or $1$.  $\alpha, \beta, \ldots$ 
are $Sp(4)$ indices and run over $1,2,3,4$.}
In section five, we then examine the effect of the walls 
and find that, in agreement with DHN,
that the dynamics of the model are controlled by $GL(2,\mathbb{Z})$. 
We conclude this section by observing that $GL(2,\mathbb{Z})$ is contained in $Sp(4,\mathbb{Z})$
so perhaps this will also lead to some progress. Lastly, we observe that a similar treatment of M-theory
can be carried out because the minisuperspace involved is just ten-dimensional Minkowski 
space and so the analog of $SL(2,\mathbb{R})$ is $SO(9,1)$ and a similar construction in 
terms of spinors can be carried out. Again this can be promoted to the conformal group
$SO(10,2)$. These last observations hint a possible octonionic viewpoint to the M-theory case.

\section{The Canonical Formalism and BKL}
\label{sec:canon}

In the canonical version of general relativity
\cite{Dirac:1958sc,Arnowitt:1962hi,DeWitt:1967yk}}, one decomposes the metric into the lapse $N$,  
the shift $N^i$ and a spatial metric 
$\gamma_{ij}$. The four-dimensional spacetime metric $g_{ab}$
is then
\beq  ds^2 = -N^2dt^2 + \gamma_{ij}(dx^i + N^idt)(dx^j + N^jdt).  \eeq
Starting from the Einstein-Hilbert action, we find that both $N$ and $N^i$ have vanishing conjugate 
momenta and 
are therefore pure gauge. The momentum conjugate to $\gamma_{ij}$ is
\beq \pi_{ij} = \frac{1}{2N}\Bigl(D_iN_j + D_jN_i - \frac{\partial \gamma_{ij}}{\partial t}\Bigr). \eeq
In this expressions and what follows,  the indices $ i,j \ldots $ are raised or lowered with respect
to $\gamma_{ij}$ and $D$ is the covariant derivative constructed from $\gamma_{ij}$.
There are also some secondary constraints. 
The first of these is  the Hamiltonian constraint ${\mathcal{H}}$.
\beq {\mathcal{H}}=-\gamma^{-1/2}\ G^{ijkl}\pi_{ij}\pi_{kl} +
 \gamma^{1/2}\ {}^{(3)}R \approx 0.\eeq
In the above expression, $\gamma = -\det{\gamma_{ij}}$, $G^{ijkl}$ is the DeWitt metric given by
\beq G^{ijkl} =  \half(\gamma^{ik}\gamma^{jl} + \gamma^{il}\gamma^{jk} - \gamma^{ij}\gamma^{kl}) \eeq
and ${}^{(3)}R$ is the Ricci scalar of the three metric $\gamma_{ij}$.
The Hamiltonian constraint is a bit like the mass-shell constraint in special relativity. 
The first term is usually referred to as the kinetic term  and is the square of the momenta
evaluated using the DeWitt metric. The DeWitt metric acts on the six components of $\pi_{ij}$
and has signature $(-++++\, +)$.
The second term is rather like a potential energy term, but as might be expected with pure gravitation, 
it is of an entirely geometric origin.
The remaining constraints are the diffeomorphism constraints, namely
\beq \chi^i = D_j\pi^{ij} \approx 0. \eeq
However they play no explicit role in what follows. 
The theory is therefore one in which the spatial metric evolves in time. 
The space of all spatial metrics quotiented by their diffeomorphisms is
termed superspace. A truncation to a simple class of spatial metrics is termed minisuperspace. 

In the BKL limit, spatial points become independent of each other, 
\cite{Lifshitz:1963ps, Belinsky:1970ew, Damour:2001sa, Damour:2002et, Henneaux:2007ej, Belinski:2017fas}. 
This is a result of 
a considerable simplification of the potential term. The kinetic term can best be understood by making 
an Iwasawa decomposition of the three metric. Write $\gamma_{ij} = (P^TDP)_{ij}$ with $P$ 
being upper unitriangular
and $D$ being diagonal and given by
\beq D = {\rm diag}\ ( e^{-2\beta^1}, \ e^{-2\beta^2},\ e^{-2\beta^3} ).\eeq
The BKL limit has the effect of making $P$ become time independent so that it
 disappears from the kinetic term. Assuming the potential term vanishes, 
the resultant Hamiltonian constraint would be
\beq {\mathcal{H}}= G_{ij}\frac{\partial \beta^i}{\partial t}\frac{\partial\beta^j}{\partial t}. \eeq 
The restricted DeWitt metric $G_{ij}$ for the diagonal piece is three dimensional and is 
\beq G_{ij} = \bpm 0 & & -1 & & -1 \\ -1 & & 0 & & -1 \\ -1 & & -1 & & 0 \epm. \eeq 
Thus if the Hamiltonian constraint were just the kinetic term, the complete spacetime
would be described by null geodesics in the flat geometry given by $G_{ij}$.
Each point in space would then be described by the Kasner metric.
\beq ds^2 = - dt^2  + \vert t\vert^{2p_1}d\beta^{1^2}  + \vert t\vert^{2p_2}d\beta^{2^2} 
+ \vert t\vert^{2p_3}d\beta^{3^2} \eeq
with $p_1+p_2+p_3=p_1^2+p_2^2+p_3^2=1$. Hence two of the $p_i$ are positive and one negative. 
There is a curvature singularity at $t=0$, so if $t\ge 0$ the geometry is an initial singularity and if
$t\le 0$ it is a final singularity.
 
Fortunately, the potential becomes very simple in the BKL limit. It is either zero or positive infinity.
The discontinuities occur at the impenetrable walls $\beta^1=0,\beta^3=\beta^2$ and $\beta^1=\beta^2.$
A geodesic will be specularly reflected by these walls.
The result is to restrict the
range of $\beta^i$ to the region bounded by $\beta^1\ge 0,\ \beta^3\ge \beta^2$ and $\beta^2-\beta^1\ge 0$.
Thus the evolution of the classical spacetime is a piecewise null geodesic in this minisuperspace
but with reflection when it meets the walls. In terms of the spacetime, 
it is a sequence of Kasner universes.

To make our treatment the more transparent, one can find a coordinate transformation turning $\beta^i$
into the usual coordinates $X^0, X^1, X^2$ in three-dimensional Minkowski spacetime.
\beq X^0 =\frac{1}{\sqrt{2}}(2\beta^1+\beta^2+\beta^3), \label{eq:xo}\eeq
\beq X^1 = \sqrt{2}\beta^1 \label{eq:x1}, \eeq
\beq X^2 = \frac{1}{\sqrt{2}}(\beta^3-\beta^2) \label{eq:x2}.\eeq
Then the truncated DeWitt metric becomes
\beq {\rm diag}(-1,1,1). \eeq

\section{$SL(2,\mathbb R)$}
\label{sec:sl2}

As we saw in the previous section, solutions of the four-dimensional Einstein equations in the 
BKL limit are locally null geodesics in
a flat three-dimensional minisuperspace of signature $(-+\, +)$.
Therefore a Bianchi I universe is equivalent to free massless particle motion in Minkowski space.
The null geodesics in this version of minisuperspace in the coordinates of (\ref{eq:xo})-(\ref{eq:x2}) are
\beq X^0 = \eta^2 s + \hat X^0, \eeq
\beq X^1 = \eta^2 s \cos\theta + \hat X^1,  \eeq
\beq X^2 = \eta^2 s \sin\theta + \hat X^2. \eeq
In the above, $s$ is an affine parameter, $\theta$ is the angle from the $X^1$-axis in the 
spatial $X^1$-$X^2$
plane, $\eta^2$ is the analogue of energy and finally $\hat X^0, \hat X^1$ and $\hat X^2$ give the position
when $s=0$.
The momentum $P^a$ is given by differentiation of $X^a$
with respect to $s$ and is null. In components
\beq P^0 = \eta^2, \eeq
\beq P^1 = \eta^2 \cos\theta, \eeq
\beq P^2 = \eta^2 \sin\theta. \eeq
In general, the specification of a null geodesic in three-dimensions requires four parameters, 
whereas our
description involves six quantities $s,\theta,\eta,\hat X^0, \hat X^1$ and $\hat X^2$. Two 
of these quantities are consequently redundant, rather like gauge parameters.  We will now outline
an approach which allows a more compact presentation.
The Clifford algebra in $2$+$1$ dimensions is given by a set of gamma matrices $\gamma^a{}{^{}_A}{}^B$
defined by
\beq \{\gamma^a{}{^{}_\sA}{}^\sB,\gamma^b{}{^{}_\sB}{}^\sC \} = 2\eta^{ab} \delta_\sA^\sC, \eeq
$\eta^{ab}$ being diagonal $(-++)$ and is used to raise or lower vector indices.
Whilst not absolutely necessary, it lends clarity to have in mind an explicit realisation of the Clifford
algebra. We will employ the following representation in our calculations, frequently giving results
in terms of the components of various geometrical quantities.
 We will use for the gamma matrices 
\beq \gamma^0{}_{\sA}{}^{\sB} = \bpm  0 & -1 \\ 1 & 0  \epm, \label{eq:gamma0}\eeq
\beq \gamma^1{}_{\sA}{}^{\sB}= \bpm  0 & 1 \\ 1 & 0 \epm, \label{eq:gamma1}\eeq
\beq \gamma^2{}_{\sA}{}^{\sB} = \bpm 1 & 0 \\ 0 & -1 \epm. \label{eq:gamma2} \eeq
Spinor indices are raised using left multiplication by the charge conjugation matrix, or perhaps 
more properly, the symplectic form
\beq \epsilon^{\sA \sB}  = \bpm 0 & 1 \\ -1 & 0 \epm. \eeq
Similarly, spinor indices are lowered using right multiplication by
\beq \epsilon_{\sA \sB} = \bpm 0 & 1 \\ -1 & 0 \epm. \eeq

$SL(2,\mathbb R)$ is the double cover of $SO(2,1)$. A well-known consequence is that
one can define the coordinates in terms of spinors \cite{Penrose:1967wn,Penrose:1972ia,Kugo:1982bn} by
\beq X^{\sA \sB} = \gamma_a{}^{\sA \sB} X^a. \eeq
Using our conventions, the components of $X^{\sA \sB}$ are 
\beq X^{\sA \sB} = \bpm -X^0+X^1 & -X^2 \\ -X^2 & -X^0-X^1 \epm. \eeq
A Lorentz invariant quantity is $\det X^{\sA \sB}$ since
\beq \det X^{\sA \sB} = (X^0)^2 - (X^1)^2 - (X^2)^2. \eeq
If $\det X^{\sA \sB}$ vanishes, the rank of $X^{\sA \sB}$ is one unless $X^{\sA \sB}$ is identically zero.
A null vector can then be written in terms of a single spinor. The momentum
spinor $\pi^\sA$ is defined by
\beq P^a = \gamma^a{}{^{}_{\sA \sB}}\pi^\sA\pi^\sB. \label{eq:momentum} \eeq
$\pi^\sA$ is therefore determined by $P^a$ upto an overall sign ambiguity. 
Thus
\beq \pi^0 = \pm\,\eta\,\sin\half\theta, \ \  \ \ \ \pi^1 = \pm\,\eta\,\cos\half\theta. \eeq
For convenience, we will take the plus signs for both components of $\pi^A$ unless otherwise stated.

The angular momentum about the origin is
\beq M^{ab}=2X^{[a}P^{b]}. \eeq
In components
\beq M^{01} = \eta^2(\hat X^0\,\cos\theta - \hat X^1), \eeq 
\beq M^{02} = \eta^2(\hat X^0\,\sin\theta - \hat X^2), \eeq
\beq M^{12} = \eta^2(\hat X^1\,\sin\theta - \hat X^2\,\cos\theta). \eeq
$M^{ab}$ can be turned into the angular momentum spinor $M^{AB}$
by \footnote{\ \ \ $\gamma_{ab}=\half(\gamma_a\gamma_b
-\gamma_b\gamma_a)$.} 
\beq M^{\sA \sB} = \half\,M^{ab}\gamma_{ab}{}^{\sA \sB}.\eeq
Again, in terms of components
\beq M^{00} = \eta^2((-\hat X^0 + \hat X^1)\sin\theta + \hat X^2(1-\cos\theta)), \eeq
\beq M^{01}=M^{10}= \eta^2(-\hat X^0\cos\theta + \hat X^1), \eeq
\beq M^{11} = \eta^2((\hat X^0 + \hat X^1)\sin\theta - \hat X^2(1+\cos\theta)). \eeq
One now finds that there is a spinor $\omega^A$ that contains all the information about the 
location of the trajectory since
\beq M^{\sA \sB} = \pi^\sA\omega^\sB + \omega^\sA \pi^\sB. \eeq
We now find that
\beq \omega^\sA = -\hat X^{\sA\sB}\pi{^{}_\sB}, \eeq
or in component form
\beq \omega^0  = \eta\,((-\hat X^0 +\hat X^1)\cos\half\theta + \hat X^2\sin\half\theta ), \eeq
\beq \omega^1  = \eta\,((\hat X^0 + \hat X^1)\sin\half\theta - \hat X^2\cos\half\theta).\eeq	
Note that if $\hat X^a$ is translated along the null geodesic, $\omega^\sA$ is invariant.

The spinorial derivative operator $\nabla{^{}_{\sA \sB}}$ is defined by 
\beq \nabla{^{}_{\sA \sB}} = \gamma^a{}{^{}_{\sA \sB}}\nabla_a. \eeq
In our flat minisuperspace, it takes the explicit form
\beq \nabla{^{}_{\sA \sB}} = \bpm \partial_0 - \partial_1 & \partial_2 \\
 \partial_2 & \partial_0 + \partial_1 
\epm.  \eeq
If we put a hat over the derivative operator, instead taking the derivative with respect to $X$
it is taken with respect to $\hat X$. We then find that $\omega^{\sA}$ obeys the twistor equation
\cite{Penrose:1967wn,Penrose:1972ia}
\beq \hat\nabla^{(\sA \sB}\omega^{\sC)} = 0. \eeq
Furthermore, $\omega^\sA$ contains all the information about the trajectory since the momentum
spinor can be found by taking its derivative
\beq \pi_\sA = \tfrac{1}{3}\hat\nabla{^{}_{\sA \sB}}\omega^B.\label{eq:defpi}\eeq

Now consider a vector $W^a$ defined by
\beq W^a = \gamma^a{}{^{}_{\sA \sB}}\,\omega^\sA\omega^\sB. \eeq
In components
\beq W^0 = \eta^2((\hat X^{0^2} + \hat X^{1^2} + \hat X^{2^2})  - 2\hat X^0 \hat X^1\cos\theta
-2\hat X^0 \hat X^2 \sin\theta), \eeq
\
\beq W^1 = \eta^2(2\hat X^0 \hat X^1 + (-\hat X^{0^2} - \hat X^{1^2} + X^{2^2})\cos\theta
-2\hat X^0 \hat X^1 \sin\theta), \eeq
\
\beq W^2 = \eta^2(2\hat X^0 \hat X^2 - 2\hat X^1 \hat X^2\cos\theta + (-\hat X^{0^2} + \hat X^{1^2} -
 X^{2^2})\sin\theta).\eeq
$W^a$ obeys the conformal Killing equation
\beq \hat\nabla_aW_b + \hat\nabla_bW_a - \tfrac{2}{3}\eta_{ab}\hat\nabla_cW^c = 0. \eeq
Additionally, we find
\beq \hat\nabla_aW^a = 6\pi_\sA\omega^\sA. \eeq
These last observations hint that we should investigate conformal transformations of our system.
Suppose we make a conformal transformation on the metric
\beq g_{ab} \mapsto \tilde g_{ab} = \Omega^2 g_{ab}. \eeq
The covariant derivative of spinor $\psi_{\sA}$ makes a corresponding transformation
\beq \nabla{^{}_{\sA \sB}}\psi{^{}_\sC} \mapsto \tilde \nabla{^{}_{\sA \sB}}\psi{^{}_\sC} 
=  \nabla{^{}_{\sA \sB}}\psi{^{}_\sC} - 
\tfrac{3}{2}\Upsilon{^{}_{\sC(\sA}}\psi{^{}_{\sB)}}
-\tfrac{3}{2}\epsilon{^{}_{\sC(\sA}}\Upsilon{^{}_{\sB)\sD}}\psi^\sD \eeq
where
\beq \Upsilon{^{}_{\sA \sB}} =\tfrac{1}{3} \nabla{^{}_{\sA \sB}}\ln \Omega. \eeq
We can now exhibit how $\pi_\sA$ and $\omega^\sA$ transform under translations and 
under conformal transformations.
Suppose we make the translation $\hat X^a \mapsto \hat X^a + \bar X^a,$
then $\pi^\sA$ is invariant but $\omega^\sA$ transforms non-trivially as
\beq \pi{^{}_\sA} \mapsto \pi{^{}_\sA}, \ \ \ \ \ \ \omega^\sA \mapsto \omega^\sA 
- \bar X^{\sA \sB}\pi{^{}_\sB}. \eeq
Under conformal transformations it is $\omega^\sA$ that is invariant and $\pi{^{}_\sA}$ transforms 
non-trivially
\beq \pi{^{}_\sA} \mapsto  \pi{^{}_\sA} + \Upsilon{^{}_{\sA \sB}}\omega^\sB, \ \ \ \ \ \ \omega^\sA \mapsto \omega^\sA. \eeq
The transformation of $\pi{^{}_\sA}$ is most easily seen to be a consequence of (\ref{eq:defpi}).
Lastly, we note that both $\pi^A$ and $\omega_A$ are Majorana spinors of $SO(2,1)$ since
they are real and thus charge conjugation and Dirac conjugation are equal upto an irrelevant overall sign.
\footnote{\ I would like to thnak Gary Gibbons for this observation.}

\section{$Sp(4,\mathbb R)$}
\label{conformal}
The symplectic group $Sp(4,\mathbb R)$ is defined to be the set of transformations $M$
preserving the symplectic form $\Omega$. Thus
\beq M^T\, J\, M = J \eeq
with
\beq J = \bpm 0 & & \mathbb{1} \\ - \mathbb{1}& & 0 \epm \label{eq:J} \eeq
and $\mathbb{1}$\   being the unit $2 \times 2$ matrix.
Let $M$ be written in block $2 \times 2$ form as
\beq M = \bpm A & & B \\ C & & D \epm \label{eq:symp} \eeq
then
\beq A^TC-C^TA=0, \eeq
\beq B^TD-D^TB=0, \eeq
\beq A^TD-C^TB=\mathbb{1}. \eeq 
These relations result in the generators of the algebra being of the form
\beq \bpm {-X^T} & & 0 \\ 0 & &  X \epm,\ \  \bpm 0 & & Y \\ 0 & & 0 \epm
\ {\rm and} \  \bpm 0 & & 0 \\ Z & & 0 \epm\eeq
with $Y$ and $Z$ being symmetric and $X$ having no symmetry property. We see from this
that the algebra of $Sp(4,\mathbb{R})$ is ten-dimensional. $SO(3,2)$ is the conformal group of 
three-dimensional Minkowski space and $Sp(4,\mathbb{R})$ is its double cover. As such it is convenient to
write the generators of $Sp(4,\mathbb{R})$ in terms of the Clifford algebra of $SO(3,2)$. The gamma matrices
of $SO(3,2)$ are $\gamma^0$ and $\gamma^\nat$ in the timelike directions and $\gamma^1, \gamma^2$ and 
$\gamma^3$ in the spacelike directions. Just as was the case for $SO(2,1)$, an explicit representation 
is useful for clarity,
\beq \gamma^0{}^\alpha{}{^{}_\beta} = \bpm \gamma^0{}^\sA{}{^{}_\sB} & 0 \\
 0 & \gamma^0{}{^{}_\sA}{}^\sB \epm \ , \ \ \ 
\gamma^\nat{}^\alpha{}{^{}_\beta} = \bpm 0 & \epsilon^{\sA \sB} \\ \epsilon{^{}_{\sA \sB}} & 0 \epm \eeq
and 
\beq \gamma^1{}^\alpha{}{^{}_\beta} = \bpm \gamma^1{}^\sA{}{^{}_\sB} & 0
 \\ 0 & \gamma^1{}{^{}_\sA}{}^\sB \epm \ ,\ \ \ 
\gamma^2{}^\alpha{}{^{}_\beta} = \bpm \gamma^2{}^\sA{}{^{}_\sB}& 0 \\ 0 & \gamma^2{}{^{}_\sA}{}^\sB \epm
\ , \ \ \
\gamma^3{}^\alpha{}{^{}_\beta} = \bpm 0 & -\epsilon^{\sA \sB} \\ \epsilon{^{}_{\sA \sB}} & 0 \epm. \eeq
In the these expressions, the $2 \times 2$ gamma matrices are the gamma matrices of
$SO(2,1)$. Note carefully the placement of both the $SO(2,1)$ 
\footnote{$\gamma^a{}^\sA{}{^{}_\sB}$ is the transpose of $\gamma^a{}{^{}_\sA}{}^\sB$. 
From (\ref{eq:gamma0}-\ref{eq:gamma2}) we see that $\gamma^0$ is antisymmetric and $\gamma^1, \gamma^2$ are
symmetric.} and $SO(3,2)$ spinor indices. The generators of $Sp(4,\mathbb{R})$ are the antisymmetric
product of two gamma matrices. It can be shown that the symmetric part of $X$ are the Lorentz 
transformations, the antisymmetric part of X is the dilatation, $Y$ are the translations and $Z$ 
are the special conformal transformations. 

The pair of $SO(2,1)$ spinors $\omega^\sA, \pi_\sA$ can be combined into a spinor $Z^\alpha$
of the conformal group $SO(3,2)$. $Z^\alpha$ is usually referred to as a twistor and has components
\beq Z^\alpha = \bpm \omega^\sA, &\  \pi_\sA \epm. \eeq
Under
 the action of $Sp(4,\mathbb{R})$
 \beq Z^\alpha \mapsto \hat Z^\alpha = M^\alpha{}{^{}_\beta} Z^\beta. \eeq
Putting $C=0$ and therefore $A^{-1}=D^T$ in (\ref{eq:symp})  we find the Lorentz 
transformations and dilatations.
Similarly, putting $A=D=\mathbb{1},\ C=0$ generates a translation with $\hat X^0$ being shifted by
$\half(B^{00}+B^{11})$, $\hat X^1$ by $\half (-B^{00}+B^{11})$ and $\hat X^2$ by$B^{01}$.
To produce a special conformal transformation put $B=0$ and $A=D=\mathbb{1}$, then 
$C_{\sA\sB}= \Upsilon{\sA\sB}$ and
\beq \Omega = 1 - \half\gamma_a{}^{\sA\sB}C{^{}_{\sA\sB}}X^a. \eeq
The twistorial indices can be lowered to find the conjugate spinor by left 
multiplication by the symplectic form $J$ given by
(\ref{eq:J}). Similarly, they can be raised by right multiplication by $J$. 
From this we see that
\beq Z_\alpha = \bpm \pi{}{^{}_\sA}, &\ -\omega^\sA \epm. \eeq
The twistor is null in the sense that $Z^\alpha Z_\alpha = 0 $.
Each twistor is a complete description of the null geodesic in minisuperspace and therefore a 
description of the entire Kasner spacetime. In the twistorial description, we have managed to replace 
the evolution of a spacelike surface in spacetime by a point in twistor space. 

One can of course reconstruct points in minisuperspace. 
One way of doing this is to ask if two null lines in minisuperspace intersect.
Suppose we consider a second twistor 
\beq Y^\alpha = \bpm \xi^\sA, & \ \pi{^{}_\sA} \epm \eeq
then the pair of null 
lines could possibly intersect at a point $X^{\sA\sB}$ in minisuperspace. 
This requires a simultaneous  solution 
to
\beq \omega^\sA = -X^{\sA\sB}\pi{^{}_\sB} \ \ \ {\rm and} \ 
\ \ \xi^\sA = -X^{\sA\sB}\eta{^{}_\sB}.\label{eq:incid} \eeq 
contracting the first relation with $\eta{^{}_\sA}$ and the second with $\pi{^{}_\sA}$ and subtracting gives
\beq \eta{^{}_\sA}\omega^A - \pi{^{}_\sA}\xi^\sA = 0 \eeq
or more succinctly in twistorial terms
\beq Z^\alpha Y_\alpha = 0. \label{eq:incidence}\eeq
(\ref{eq:incidence}) is a necessary and sufficient condition for the two lines to intersect.
To find $X^{\sA\sB}$ requires a little algebra and we find that
\beq X^{\sA\sB} = (\xi^\sA\pi^\sB - \omega^\sA\eta^\sB)/{\pi^\sC\eta{^{}_\sC}}. \eeq

\section{Reflections}
\label{sec:reflect}

In the BKL limit, four-dimensional pure gravity is chaotic. Adding extra fields can destroy the chaos 
in which case the BKL spacetime is just the Bianchi I metric until the singularity is reached. In the
non-chaotic case, the spacetime is described by a single twistor as outlined in the previous section.
However, in the chaotic case, the situation is rather more complicated. 

As shown in \cite{Damour:2001sa,Damour:2002et,Henneaux:2007ej,Belinski:2017fas} and described in 
section two, the trajectory in 
minisuperspace is piecewise Bianchi I 
but interrupted by bounces off of three walls restricting the motion in minisuperspace to be
\beq {\it Wall\ \ I:} \ \ \ \ X^1 \ge 0, \eeq
\beq {\it Wall\ \  II:} \ \ \ \ X^2 \ge 0, \eeq
\beq {\it Wall \ \ III:} \ \ \ \ X^0 - 2X^1 - X^2 \ge 0. \eeq
Motion in minisuperspace is thus confined to the interior of a double cone of triangular cross-section,
with the  apex of the cone being at the origin, 
$X^0=X^1=X^2=0$. The spacetime singularity occurs when the trajectory in superspace reaches 
$X^0=\pm\infty$, which typcially occurs after a finite amount of proper time $t$.
Let $n^a$ be the unit spacelike vector to the walls, then the effect of a bounce is to change
both $X^a(s)$ and $P^a$ as
\beq X^a \mapsto \bar X^a = X^a - 2\,(X_b\,n^b)\,n^a,
 \ \ \ P^a \mapsto \bar P^a = P^a -2\,(P_b\,n^b)\,n^a. \eeq 
This will induce a transformation on both $\pi_\sA$ and $\omega_\sA$.

The effect in minisuperspace of Wall I is $\theta \mapsto \bar\theta=-\theta, \hat X^2 \mapsto -\hat X^2$
with other quantities invariant. Then on the spinorial quantities we find \beq \pi_0 \mapsto -\pi_1,\ \ 
\pi_1 \mapsto -\pi_0,\ \  \omega^0 \mapsto -\omega^1\ \  {\rm and} \ \ \omega^1 \mapsto -\omega^0. \eeq
The effect in minisuperspace of Wall II is $\theta \mapsto \bar\theta=\pi-\theta, \hat X^1 \mapsto -\hat X^1$
with other quantities invariant. Then on the spinorial quantities we find \beq \pi_0 {\rm\ is\ invariant},\ \ 
\pi_1 \mapsto -\pi_1,\ \  \omega^0 {\rm\ is\ invariant}\ \  {\rm and} \ \ \omega^1 \mapsto -\omega^1. \eeq
The effect in minisuperspace of Wall III is rather more complicated.
\beq X^0 \mapsto \bar X^0 = \eta^2(\tfrac{3}{2}-\cos\theta - \half\sin\theta), \eeq
\beq X^1 \mapsto \bar X^1 = \eta^2(1-\cos\theta - \sin\theta), \eeq
\beq X^2 \mapsto \bar X^2 = \eta^2(\half - \cos\theta + \half\sin\theta).\eeq
On the spinorial quantities we find \beq \pi_0 \mapsto \half\pi_0 - \tfrac{3}{2}\pi_1,\ \ 
\pi_1 \mapsto \half\pi_0 - \half\pi_1,\ \  \omega^0 \mapsto -\half\omega^0 -\half\omega^1\ \ 
 {\rm and} \ \ \omega^1 \mapsto -\tfrac{3}{2}\omega^0+\half\omega^1. \eeq
In twistor space one can summarise these relations simply as
\beq Z^\alpha \mapsto \bar Z^\alpha = S_{(i)}\,{}^\alpha{\,}_\beta\, Z^\beta \eeq
with
\beq S_{(i)} = \bpm -\gamma_a{}^T n_{(i)}{\,}^a & & 0 \\ 0 & & \gamma_a\, n_{(i)}{\,}^a \epm \eeq
where $n_{(i)}{}^a$ is the unit normal to the $i^{th}$ wall and $\gamma^a$ being the $SO(2,1)$ gamma matrices
of (\ref{eq:gamma0}) - (\ref{eq:gamma2}). Note that these are of the form of an improper 
Lorentz transformation and dilatation.  Also, as expected from reflections, $S_{(i)}^2=1$. 

We can now find the Coxeter group of these reflections.
\beq (S_{(1)}S_{(2)})^2 = -1, \ \  (S_{(2)}S_{(3)})^3 = 1, \ \ (S_{(1)}S_{(3)})^\infty = 1.\eeq
In the last case, infinity is a convention that is used to denote that no positive integer power
leads to unity. 
Recalling that $\pi_\sA$ and $\omega^\sA$ appear 
quadratically in  $P^a$ and $X^a$, the
fact that $(S_{(1)}S_{(2)})^2=-1$ is equivalent to the identity when acting on $P^a$ and $X^a$. 
The Coxeter group relevant to the DHN picture is therefore the triangle group $I(2,3,\infty)$.

In \cite{Damour:2001sa,Damour:2002et,Henneaux:2007ej,Belinski:2017fas}, the trajectory in minisuperspace is
illustrated by its projection into the Lobachevsky plane. Let $w$ be a complex coordinate in the
Lobachevsky plane yielding as its metric
\beq \frac{dw d\bar w}{ ({\rm Im}\,w)^2}\ . \eeq
The transformation from the $X^a$ to the complex $w$-plane is rather complicated and is
spelt out in the appendix. 
Motion is restricted to the region outside the unit circle in
complex $w$ space and also between $ {\rm Re}\, w = 0$ and $ {\rm Re}\, w=\half $. One quickly recognises 
this as half the fundamental region of $SL(2,\mathbb{Z})$. In fact, this is the 
the fundamental region of $PGL(2,\mathbb{Z})$.  

Reflections generate
the  Cartan matrix
\beq A_{ij} = 2\frac{n_{(i)}{}^a n_{(j)}{}_a}{n_{(i)}{}^a n_{(i)}{}_a}. \eeq
In our case, we can evaluate $A_{ij}$ to give
\beq \bpm 2 & 0 & -2 \\ 0 & 2 & -1 \\ -2 & -1 & 0 \epm. \eeq
$A_{ij}$ is the Cartan matrix of the hyperbolic Kac-Moody algebra $A_1{}^{++}$.
Thus $PGL(2,\mathbb{Z})$ is the Weyl group of $A_1{}^{++}$. 
The twistor approach reproduces this result which was originally found using the minisuperspace picture.

\section{Conclusions and Speculations}
\label{sec:conc}
What we have achieved is a new description of the BKL phenomenon in terms of twistors. The twistors 
encode all of the phase space of the BKL limit. A point, or if there is chaos a collection of 
points, in twistor space corresponds to
an entire spacetime. This is in contrast to the minisuperspace description where a spacetime is a
trajectory. If we consider now the conjecture of DHN, namely that BKL
exposes the fundamental degrees of freedom of gravitation, then this alternative might allow
a little more progress to be made.  One challenge is to reproduce the semi-classical
picture of gravitation. We have introduced some discreteness in the form of the
triangle group acting on the twistors. In the DHN picture $SL(2,\mathbb{R})$
morphs into the discrete group $GL(2,\mathbb{Z})$ and it is this observation that leads  to the
speculation that the fundamental nature of gravity is thereby revealed.
Whilst that did not quite work for reasons that are unknown, it perhaps seems unnatural in the twistor
approach to only make part of $Sp(4,\mathbb{R})$ discrete. Feingold and Frenkel \cite{Feingold:1983}
showed how to extend
this picture to $Sp(4,\mathbb{Z})$. If this holds is the twistorial case, instead of quantisation
being about the modular forms of $GL(2,\mathbb{Z})$ \cite{Kleinschmidt:2009cv, Perry:2021mch}
it will be about the Siegel modular forms of genus
two. 

It is possible that Bianchi I is not the only regime that can be described in this way. For example
Bianchi VIII and IX universes lead to a similar type of chaos but with different Coxeter groups.
Different starting points therefore are potentially interesting and should therefore be examined
to see if similar results can be obtained. Perhaps the treatment could be extended to more 
complicated spacetimes too.

There is a possible connection with moonshine as all of the
sporadic discrete groups can be obtained from triangle groups. In particular, twenty of the sporadic groups
can be obtained from the triangle group $I(2,3,n)$ whilst the remainder can be obtained from 
different triangle groups \cite{Wilson:2001}. $I(2,3,\infty)$ 
can be generated by two generators $S$ and $T$ obeying
$S^2=(ST)^3=\mathbb{1}$. $I(2,3,n)$ is obtained by adding the relation $T^n=\mathbb{1}$. 
It is believed that M-theory has some relation to moonshine \cite{Anagiannis:2018jqf}.

What, if anything,  does our approach have to M-theory. M-theory is based on an eleven-dimensional spacetime. 
This will result in space being ten-dimensional. In particular in the BKL limit, DHN found that M-theory 
was chaotic and the Coxeter group that controlled the chaos was $E_{10}$.
The minisuperspace in the BKL limit is ten-dimensional Minkowski space and the physical spacetime is again
described by a null vector in minisuperspace. The Lorentz group of the minisuperspace is SO(9,1) 
which is contained in the 
conformal group $SO(10,2)$ in way that is precisely the same as the three-dimensional case. 
However, $SO(10,2)$ is not related to a classical symplectic group. 
The appropriate gamma matrices again 
have the property that the momentum can be written in terms of a spinor, exactly as in 
(\ref{eq:momentum}). 
As is well-known, this latter property only holds in dimensions 
$3,4,6\ {\rm and}\ 10$ when the signature is Lorentzian and so only spacetimes
of dimension $3,4,7\ {\rm and}\ 11$ can easily be described this way.
These numbers are of course related to the existence of the four division algebras. 
That motivates another possible way to look at the M-theory example. A null vector can be
represented by an element of of $SL(2,\mathbb{O})$  where $\mathbb{O}$ are the octonions.  The 
Lorentz algebra of $SO(9,1)$ is 
then the direct sum of $L$ and $G_2$ where $L$ is the space of octonionic $ 2 \times 2 $ matrices
and $\{\Lambda \in L\ \vert \ {\rm Re \ (tr}\, \Lambda) = 0 \}$,
\cite{Kugo:1982bn,Sudbery:1984,Hitchin:2018}. The conformal algebra
is now $Sp(4,\mathbb{O})$, a precise definition of which was found by 
Sudbery and Chung  \cite{Chung:1987in}. It may be possible that the similarity between
the case discussed in this paper using the reals and the M-theory version using the octonions 
can be fruitfully exploited to describe M-theory.
The more so that it is known that $E_{10}$ is related to the octonions, 
\cite{Feingold:2008ih, Baez:2014}. 
 Baez \cite{Baez:2001dm} has described
octonions as "the crazy old uncle that nobody lets out of the attic.\rq\rq\
The idea that octonions are somehow related to physics is an old one and although fraught with difficulties
it is  nevertheless a stimulating and intriguing possibility.

An examination of some of these issues will be contained in future publications.
 
\acknowledgments

I would like to thank the STFC for financial support under grant ST/L000415/1.
I would also like to thank Alex Feingold, Gary Gibbons, Mahdi Godazgar, Isaak Khalatnikov, 
Axel Kleinschmidt, 
Hermann Nicolai, 
Roger Penrose and Chris Pope for interesting
discussions and for providing insights into the matters discuused here.

\vskip 4cm

\section{Appendix}

Here we summarise the coordinate transformations that enable us to go from three-dimensional Minkowski
minisuperspace to the complex $w$-plane. The Minkowski metric is 
\beq ds^2 = -(dX^0)^2 + (dX^1)^2 + (dX^2)^2. \eeq
Firstly, we foliate Minkowski space by a series of two-dimensional surfaces of constant negative 
curvature.
\beq X^0 = \frac{\rho}{\sqrt{12}}\Bigl[4\cosh r - \sinh r\Bigl(\sin\phi + \sqrt{3}\cos\phi\Bigr)\Bigr], \eeq
\beq X^1 = \frac{\rho}{\sqrt{6}}\Bigl[\cosh r - \sinh r\Bigl(\sin\phi - \sqrt{3}\cos\phi\Bigr)\Bigr], \eeq
\beq X^2 = \frac{\rho}{2}\Bigl[\sinh r\Bigl(\sqrt{3}\sin\phi - \cos\phi\Bigr)\Bigl] \eeq
yielding the line element
\beq ds^2 = -d\rho^2  + \rho^2(dr^2 + \sinh^2r d\phi^2). \eeq
To map the spatial surfaces into the complex $w$-plane set
\beq w = i\ \frac{1+\tanh\half r\ e^{-i\phi}}{1-\tanh\half r\ e^{-i\phi}} \eeq
giving the line element
\beq ds^2 = - d\rho^2 + \rho^2\ \frac{dwd\bar w}{({\rm Im}\, w)\, ^2}\ .\eeq
For further details on these coordinates, see \cite{Belinski:2017fas}.

\end{document}